\documentclass{webofc}
\usepackage[utf8]{inputenc}
\usepackage[T2A]{fontenc}

\def\ffc{F_{\rm C}}
\def\ffm{F_{\rm M}}
\def\het{^3{\rm He}}

\begin{document}
\title{Sensitivity of elastic electron scattering off the $^3$He to the nucleon form factors}
%
%

	\author{\firstname{Serge} \lastname{Bondarenko}\inst{1}\fnsep\thanks{\email{bondarenko@jinr.ru}} \and
	\firstname{Valery} \lastname{Burov}\inst{1}\fnsep\thanks{\email{burov@theor.jinr.ru}} \and
	\firstname{Sergey} \lastname{Yurev}\inst{1}\fnsep\thanks{\email{yurev@jinr.ru}}
}

\institute{BLTP, Joint Institute for Nuclear Research, Dubna, 141980, Russia 
}

\abstract{%
  Elastic electron-$^3$He scattering is studied in the relativistic impulse approximation.
  The amplitudes for the three-nucleon system -- $^3$He -- are obtained by solving the
  relativistic generalization of the Faddeev equation. The charge and magnetic form factor are
  calculated and compared with the experimental data for the momentum transfer squared up to 100 fm$^{-2}$.
  The influence of the various nucleon form factors is investigated.
}
\maketitle
\section{Introduction}
\label{intro}

The study of electron-nucleus reactions are important to obtain information on the nucleon-nucleon (NN)
interaction which is crucial for understanding the structure of the strong interactions.
At this time a lot of experimental data are known for the reaction cross sections and polarization observables.
The intermediate range of energies in these reactions is mainly interesting where the nonrelativistic
description based on the potential or the nonrelativistic meson-nucleon models does not already work.
From another hand the Quantum Chromodynamics which operates with quark and gluon degrees of freedom
does not also give an appropriate description.

The planned experiments, such, for example, as Jefferson Lab Experiment E1210103 at 12 GeV, require
to take into account the relativistic treatment of the nuclear systems.

One of the promising approaches is the covariant formalism based on the Bethe-Salpeter equation for
two nucleons. This approach operates with relativistic meson-nucleon degrees of freedom.
There are a lot of investigations of the two-nucleon systems and their reactions~(see, \cite{TjonZuilhof,HummelTjon}).

The calculations with the covariant separable kernels of NN interaction are also widely used.
The results of the elastic electron-deuteron scattering investigations can be found
in~\cite{RupTjon,Bondarenko:elastic,Bondarenko:2002zz}.

To study the three-particle systems in quantum mechanics, the Faddeev equation is usually
used~\cite{vanFaassen:1986wk,sitenko}.
This equation describes these systems with a pair potential of any kind as bound and scattering states.
In the case of a system of three relativistic particles,
the relativistic generalization of the Faddeev equation
can be applied~\cite{Stadler:1997iu,Rupp:1987cw,Bondarenko:2015kma,Bondarenko:2017ibt,SPD}.
In this case the relativistic two-particles T matrix is taken as a solution
of the Bethe-Salpeter equation (we call such three-particle equation as Bethe-Salpeter-Faddeev equation, BSF)
or the corresponding covariant spectator (or Gross) equation.

To solve the BSF equation, it is necessary to know the potential of the nucleon-nucleon interaction
in the explicit form. To simplify the calculations, the separable kernels of NN interactions can be
used~\cite{Rupp:1987cw,Bondarenko:2015kma,Bondarenko:2017ibt,SPD}.
In recent works, we solved the BSF equation
for the one-rank separable kernels of NN interactions and took into account the two-nucleon states
with total angular momentum $j=0,1$~\cite{Bondarenko:2017mef}. For the sake of simplicity,
we considered the nucleon propagators
in the scalar-particle form while the spin-isospin dependence was treated by applying the recoupling
matrix~\cite{SPD}.
The form factors of the potential were the relativistic Yamaguchi
functions~\cite{Yamaguchi:1954mp}.

The relativistic calculations for elastic electron scattering off the $^3$He were considered in
several papers~\cite{Pinto:2009jz,Rupp:1987cw}.
The purpose of this work is to study the sensitivity of unpolarized elastic electron scattering off the $^3$He
to the nucleon electromagnetic (EM) form factors.
The vertex functions obtained in previous works~\cite{Rupp:1987cw,Bondarenko:2015kma}
were used in the calculation of the elastic charge $F_C$ and magnetic $F_M$ form factors of the $^3$He.
We consider various combinations of the spin, isospin and momenta
of the three nucleons under their permutations, to satisfy the Pauli principle
(the $^3$He vertex function must be antisymmetric with respect to permutation of any pair of particles).

In present calculations we consider only main $S$ partial wave states ($^1S_0$ and $^3S_1$) of the $^3$He.
The influence of the dipole fit, vector-dominance and relativistic harmonic oscillator models for
the nucleon EM form factors is investigated.

The paper is organized as follows: in Sec. 2 the expressions for the $^3$He form factors are given,
in Sec. 3 the calculations and results are discussed and finally the conclusion is given.

\section{Form factors of a three-nucleon system}
\label{sec-2}

As a system with one-half spin the electromagnetic current of the $^3$He can be parameterized by
only two elastic form factors: charge (electric) $\ffc$ and magnetic $\ffm$ (see for example,~\cite{sitenko}).
In calculations we use the straightforward relativistic generalization of the nonrelativistic expressions
for form factors for the $^3$He and they have the following form~\cite{schiff,Rupp:1987cw}:
\begin{eqnarray}
	&&2\ffc =  (2F^p_{\rm C} + F^n_{\rm C} )F_1  - \frac23(F^p_{\rm C} - F^n_{\rm C})F_2, 
	\label{F_He_ch}
	\\
	&&\mu(\het)F_{\rm M} =   F^n_{\rm M} F_1  + \frac23(F^n_{\rm M} + F^p_{\rm M})F_2, 
\nonumber
\end{eqnarray}
where $F^{p,n}_{\rm C}$ is the charge form factor of the proton and neutron, respectively,
$F^{p,n}_{\rm M}$ is the magnetic form factor of the proton and neutron, respectively,
$\mu(\het)$ is magnetic moment of the $^3$He. The $\het$ form factors are normalized to unity at
zero momentum transfer while the nucleon ones have the following normalization:
$F^p_{\rm C}(0)=F^p_{\rm M}(0)/\mu_p=F^n_{\rm M}(0)/\mu_n=1$ and $F^n_{\rm C}(0)=0$.

The functions 
$F_1$ and $F_2$ can be expressed in terms of the vertex functions of the three-nucleon system.
In the relativistic case, they have the form:
\begin{eqnarray}
	F_{1}(Q^2) =  \int d^4{p}  \int d^4{q}\,  G_1 G_2 G_3 G'_3 
	\Psi^*_{S}(p,q)\Psi_{S}(p,q'),
	\label{F1}
\label{f12}
\end{eqnarray} 
\begin{eqnarray}
	F_{2}(Q^2) = -3 \int d^4 {p}  \int d^4 {q}\, G_1 G_2 G_3 G'_3 
	\Psi^*_{S}(p,q)\Psi_{S'}(p,q'),
	\label{F2}
\nonumber
\end{eqnarray}
where $Q$ is the momentum transfer,
$p$ and $q, q'=q-\frac23Q$ are the Jacobi momenta of the three-nucleon system and
the nucleon propagators are:
\begin{eqnarray}
	G_1 =  [(\frac13 \sqrt s + i(\frac12q_4 + p_4 ))^2  - p^2 -  \frac14q^2 -  \mathbf{p} \cdot \mathbf{q} -m^2]^{-1},
\label{G123}
\end{eqnarray} 
\begin{eqnarray}
	G_2 =  [(\frac13 \sqrt s + i(\frac12q_4 - p_4 ))^2  - p^2 -  \frac14q^2 +  \mathbf{p} \cdot \mathbf{q} -m^2]^{-1},
\nonumber
\end{eqnarray} 
\begin{eqnarray}
	G_3 = [(\frac13 \sqrt s - iq_4 )^2  -  q^2 -m^2]^{-1} ,
\nonumber
\end{eqnarray} 
\begin{eqnarray}
	G'_3 = [(\frac13 \sqrt s - iq_4 )^2  -  q^2 -\frac49Q^2 + \frac43 \mathbf{q} \cdot \mathbf{Q}-m^2]^{-1},
\nonumber
\end{eqnarray}
here $m$ is the nucleon mass.

Since the solution of the BSF was found in the three-nucleon c.m.s. and the $\het$ form factors are
calculated in the Breit system, the Lorentz boost transformation for the internal momenta should be done.
The $q,q'$ are defined in the corresponding c.m.s of the initial and final $\het$, respectively,
while $Q$ -- in the Breit system. The Lorentz transformation mixes the $q_0$, $q_z$ and $Q_z$
components of the momenta such that after the Wick rotation $q_0 \to iq_4$ the $q, q'$ momentum components
become complex-valued. Since the $\het$ vertex function was obtained only for the real values of $q_4$,
it is almost impossible to apply the Lorentz transformation for the $q'$. Fortunately, since
the characteristic parameter of the transformation $\eta=-Q^2/4s$ is equal to only 0.12
at $-Q^2=100$ fm$^{-2}$, we can perform a Taylor series expansion of the expression for $q'$ near
the $q$ value and get: $q'=(q_0,\mathbf{q} - \frac23 \mathbf{Q})$. In this case the arguments
of the final $\het$ vertex function become
$(q'_4,|\mathbf{ q}'|) = (q_4,|\mathbf{q} - \frac23 \mathbf{Q}|)$.

The symmetric $\Psi_{S}$ and mixed-symmetry $\Psi_{S'}$ functions can be expressed~\cite{schiff,Rupp:1987cw}
in terms of the vertex functions found as a solution of the BSF equation in~\cite{Rupp:1987cw,SPD}.
Considering only $S$ partial-wave states in the $\het$ these functions have the following form~\cite{Rupp:1987cw}:
\begin{eqnarray}
&& \Psi_{S}(1) = A(1) + A(2) + A(3),\\
&& \Psi_{S'}(1) = \frac12(B(3) + B(2) - 2B(1)),
\label{4func1}
\end{eqnarray}
where
\begin{eqnarray}
&&  A(i)  = u_s(i) -  u_t(i),\\
&&  B(i)  = u_s(i) +  u_t(i).
\label{4func2}
\end{eqnarray}
Here $i$ is the number of particle and $u_s$ and  $u_t$ are the functions of the $^1S_0$ and $^3S_1$ states,
respectively. The expressions for $u_{s,t}$ related to the the BSF equation solutions $\Phi_{s,t}$ as follows
($a=(^1S_0,^3S_1)$):
\begin{eqnarray}
u_a(p,q) = g_a(p)\tau^{a}(q,s)\Phi_a(q),
\end{eqnarray}
which is a consequence of using the kernel of NN interaction in a separable form.

\section{Calculations and results}
\begin{figure}[ht]
  \centering
			\includegraphics[width=0.6\linewidth,angle=-90]{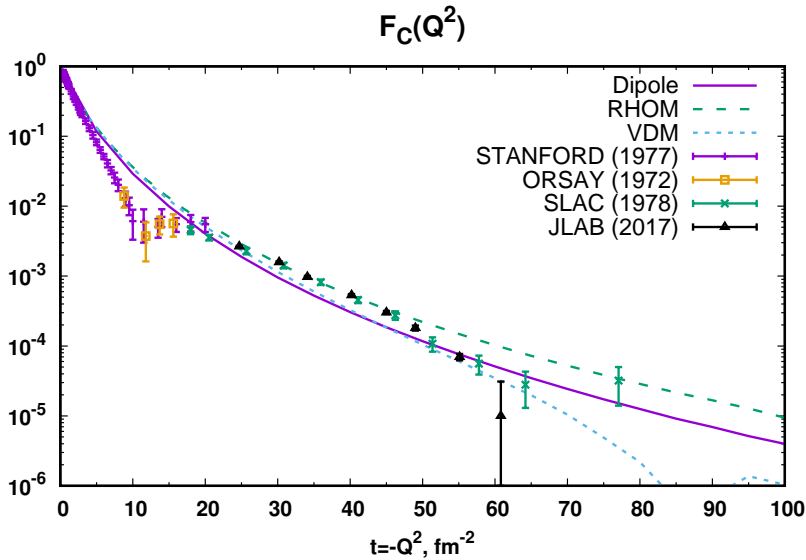}
	\caption{(Color online) The charge form factor of $^3$He. Solid line is the result of using
          the Dipole fit, long dashed line -- with RHOM, and short dashed line with VDM.
          The experimental data STANFORD~(1977) are from~\cite{stanford1977}, ORSAY~(1972) --~\cite{orsay1972},
          SLAC~(1978) --~\cite{slac1978}, JLAB~(2017) --~\cite{jlab2017}}
	\label{ff1}
\end{figure}

\begin{figure}[ht]
	\center{
		\begin{tabular}{cc}    
			\includegraphics[width=0.6\linewidth,angle=-90]{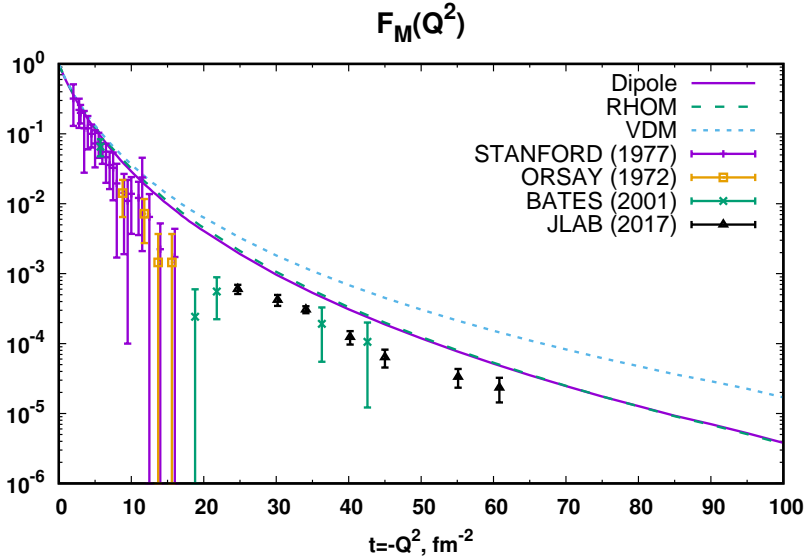}
		\end{tabular}
	}
	\caption{(Color online) The magnetic form factor of $^3$He. Notations for the results as in Fig.~\ref{ff1}.
          The experimental data STANFORD~(1977) are from~\cite{stanford1977}, ORSAY~(1972) --~\cite{orsay1972},
          BATES~(2001) --~\cite{bates2001}, JLAB~(2016) --~\cite{jlab2017}}
	\label{ff2}
\end{figure}

To calculate functions $F_{1,2}$ we used the analytic expressions for $g_a(p),\tau^{a}(q,s)$ and
interpolation of the numerical solutions for functions $\Phi_a(q)$ which were obtained by solving
the BSF system of homogeneous integral equations by means of the Gaussian quadratures
(see details in~\cite{Bondarenko:2015kma}).
The Vegas algorithm of Monte-Carlo integrator was used to perform a multiple integration in Eq.~(\ref{f12}).

To study the influence of the nucleon EM form factor, three models were considered:
the well known dipole fit (DIPOLE), the vector-dominance model (VDM) and
the relativistic harmonic
oscillator model (RHOM))~\cite{Bondarenko:elastic}.

In Fig.~\ref{ff1} the charge form factor $\ffc$ is shown for the transfer momentum squared region up to 100 fm$^{-2}$.
It is seen that the obtained results describes well the experimental data only at low 
and in the medium region of momentum transfer squared $t=-Q^2$.
The difference between calculations and the experimental data becomes
larger with increasing the transfer momentum and it depends crucially of used nucleon form factor model.
The difference between calculations themselves becomes larger with increasing the transfer momentum and
reaches the half of order at $-Q^2$=100 fm$^{-2}$. It should be stressed here that only VDM gives the interference
minimum in the $\ffc$ observed at the experiment which, however, shifted a lot to the region of high $Q^2$.

In Fig.~\ref{ff2} the magnetic form factor $\ffm$ is shown for the transfer momentum squared region up to 100 fm$^{-2}$.
As in the case of $\ffc$, the results describe the experimental data at low and partially
in the medium region of momentum transfer squared $Q^2$. All three models does not give the interference
minimum (opposite to ${\ffc}$ results), which is observed in the data at $Q^2$ about 16-18 fm$^{-2}$,
and systematically overestimated the experimental data starting from 10 fm$^{-2}$.
The DIPOLE and RHOM results practically coincides up to the $-Q^2$ = 100 fm$^{-2}$ while the VDM curve goes
above them and the difference reaches about half of the order of magnitude at $-Q^2=60$ fm$^{-2}$.

We consider the difference between results of the calculations and the experimental data as an indication
that the simple one-rank separable kernel should be improved and the multirank kernel should be considered.
We see also the rather large influence of the EM nucleon form factors to the $\het$ elastic form factors.

\section{Conclusion}
\label{summ}

In the paper the solutions of the BSF equation for the $\het$ have been used to
calculate the elastic form factors. The expressions for the form factors have been obtained by
the straightforward relativistic generalization of the nonrelativistic expressions.
The multiple integration have been performed by means of the Monte Carlo algorithms.
The influence of the various EM nucleon form factors has been investigated which has been found
to be large.

The difference between calculations as well as overestimation of the experimental data
are, supposedly, due to the simple model of the NN interaction (the separable one-rank).
Hopefully, the increasing the rank of the separability can improve the results.

This work was partially supported by the Russian Foundation for Basic Research grants {\it N} 16-02-00898
and {\it N} 18-32-00278.

\end{document}